# Toward a Knowledge-based Personalised Recommender System for Mobile App Development

**Bilal Abu-Salih**
The University of Jordan, Amman, Jordan
b.abusalih@ju.edu.jo

**Hamad Alsawalqah**
The University of Jordan, Amman, Jordan
h.sawalqah@ju.edu.jo

**Basima Elshqeirat**
The University of Jordan, Amman, Jordan
b.shoqurat@ju.edu.jo

**Tomayess Issa**
Curtin University, Perth, Australia
Tomayess.Issa@cbs.curtin.edu.au

**Pornpit Wongthongtham**
The University of Western Australia, Perth, Australia
ponnie.clark@uwa.edu.au

**Khadija Khalid Premi**
Université de Paris
dolomiza@yahoo.com

**Abstract:** Over the last few years, the arena of mobile application development has expanded considerably beyond the demand of the world's software markets. With the growing number of mobile software companies and the increasing sophistication of smartphone technology, developers have been establishing several categories of applications on dissimilar platforms. However, developers confront several challenges when undertaking mobile application projects. In particular, there is a lack of consolidated systems that can competently, promptly and efficiently provide developers with personalised services. Hence, it is essential to develop tailored systems that can recommend appropriate tools, IDEs, platforms, software components and other correlated artifacts to mobile application developers. This paper proposes a new recommender system framework comprising a robust set of techniques that are designed to provide mobile app developers with a specific platform where they can browse and search for personalised artifacts. In particular, the new recommender system framework comprises the following functions: (i) domain knowledge inference module: including various semantic web technologies and lightweight ontologies; (ii) profiling and preferencing: a new proposed time-aware multidimensional user modelling; (iii) query expansion: to improve and enhance the retrieved results by semantically augmenting users' query; and (iv) recommendation and information filtration: to make use of the aforementioned components to provide personalised services to the designated users and to answer a user's query with the minimum mismatches.

# 1 Introduction

Mobile apps have opened up vast prospects in communications and have established the means for conducting dialogues in many contexts, domains and sectors, allowing people to communicate, companies to conduct business activities, governments to provide services to their affiliated citizens, and educators to facilitate the delivery of learning materials. According to Gartner [Umuhoza and Brambilla, 2016], by 2022 enterprises will carry out 70 percent of their software interactions through mobile devices. Therefore, mobile software companies are contending to provide new and distinctive services and tools that will eventually see the emergence of mobile applications that combine the functions of a computer and a telephone and provide advanced services at various levels. However, designing and implementing a smartphone app is not a trivial undertaking. The difficulty lies in the fact that Mobile Application Development (MAD) is a sophisticated task that presents a series of challenges and requires that decisions be made throughout the development lifecycle until the point of deployment [Beyer, 2015]. However, the current development approaches used to support front and back-end tools for MAD are inadequately aligned with multi-experience requirements [Jason Wong, 2018].

The key challenges to MAD include: (1) Different mobile operating systems: the heterogeneity of specifications for the current mobile operating systems (IOS, Android, Windows Mobile, etc.) complicates the process of developing a consistent application that takes into account the hardware and software requirements of each mobile device with a different mobile operating system. (2) Different mobile development environments: the criteria for selecting the appropriate development environment for a certain mobile project domain are unclear and hard to quantify. For example, developers find it difficult to choose between the native development approach and the cross-platform development approach despite the guidance and benchmarks provided [Charkaoui and Adraoui, 2014]. (3) Peak instability of cross-platform tools and approaches: the cross-platform development ecosystem has witnessed several changes and growth in terms of the incorporated tools and technologies [Majchrzak et al., 2018]. This constant proliferation of cross-platform frameworks means that developers need to expend a significant amount of effort when selecting adequate tools which fit their expertise and the project's specifications, while simultaneously considering time and budget constraints. (4) Miscellaneous issues: the development of a smartphone application involves other related issues such as the elements of GUI design, application structure, IDE(s) selection, development cost, security and privacy, etc.

Recommender Systems (RSs) have been extensively used in various sectors [Lu 2015; Resnick and Varian 1997] , leveraging the advancements of embedded sophisticated algorithms and the profusion of supported knowledge bases [Abu-Salih et al., 2020]. Hence, RSs have brought a plethora of benefits to domains ranging from e-business [Polatidis and Georgiadis, 2013], to health informatics [Wiesner and Pfeifer 2014; Valdez et al., 2016], to social networks [Abu-Salih et al., 2020; Sun et al., 2015], to entertainment [Christensen and Schiaffino, 2011], and to many other applications [Ricci et al., 2015]. RSs in the context of software development offer services that meet developers' needs and take account of their skills and the project's specifications [Robillard et al. 2009]. These services support developers by providing a relevant and well-correlated array of prescribed solutions to their technical problems, thus saving tremendous time and effort. However, there is a lack of RSs that cater for mobile app developers. In particular, MAD entails various differences in terms of development environment and software project requirements. For example, the mechanism followed to develop and deploy a native mobile app is different from that of a regular web app. Moreover, the technicalities associated with the MAD domain require specific skills and special interface elements and development tools.

The aim of this study is to provide developers with personalised services by means of a comprehensive and time-aware knowledge-based recommender system designed to recommend and retrieve code snippets, Q&A threads, tutorials, libraries, and other external data sources and artifacts to assisting developers with their mobile app development. In particular, the new proposed RS framework comprises the following add-ons: (1) Domain knowledge inference module: including various semantic web technologies and lightweight ontologies. (2) Profiling and preferencing: a new proposed time-aware multidimensional user modelling. (3) Query expansion: to improve and enhance the retrieved results by semantically augmenting users' query. (4) Recommendation and information filtration: to make use of the aforementioned components to provide personalised services to the designated users and to answer a user's query with a minimum number of mismatches.

This paper is organised as follows: Section 2 presents the state-of-the-art review of research currently being conducted in the area of RS and its application to software development. Section 3 provides a detailed discussion of the proposed framework including all modules. Section 4 discusses the current gaps in the literature and how the framework and embedded techniques can bridge these gaps. Section 5 concludes the paper and discusses the proposed future work.

## 2 Related Works

Recently, RSs have attracted extensive attention from the research community due to the abundance of information that makes it difficult to link users with sought-after artifacts with minimal effort and time. In software engineering, the building of an operative RS is even more imperative as developers have to deal with information derived from vast and dissimilar data sources that are retrieved in different formats, such as code snippets, technical tutorials, API documentations, Q&A websites, etc. [Nguyen et al., 2018a]. It is evident that developers encounter several problems and obstacles before and during the life cycle of their software development projects [Beyer, 2015]. This section gives the technical background of this study by presenting a state-of-the-art review of key techniques which are applied when designing RSs for software development.

Katirtzis et al. in [Katirtzis et al., 2018] presented MAPO (Mining API usage Pattern from Open source repositories) system for processing source files and clustering the included API methods which are analysed to infer ranked list of API usage patterns by finding similarities with the developer context. Another attempt is MUSE [Moreno et al., 2015] which recommends to the developer certain code derived from examining source codes and cluster code snippets. Usage Pattern Miner (UP-Miner) [Wang et al., 2013] was designed to automatically mine the usage patterns of API methods from various source codes. Experiments conducted using a Microsoft codebase dataset to evaluate Up-Miner have proven its effectiveness and outweighed baseline approaches using certain metrics. Saied et al. in [Saied et al., 2015] presented a Multi-Level API Usage Patterns (MLUP) system as an approach for quarrying and inferring the co-usage relationships between various methods of the API of interest across interfering usage scenarios. They incorporate DBSCAN (Density-Based Spatial Clustering of Applications with Noise) clustering technique to assemble API methods that are commonly used coherently in software development projects. RSs for software testing are also proposed; SoTesTeR [Ibarra and Rodriguez, 2019] is a platform for recommending testing and quality assurance techniques using a content-based approach. Yadav and Dutta in [Yadav and Dutta, 2019] introduced a prioritising mechanism for software testing using a test cases clustering technique to minimise the time and effort required to conduct regression testing for software maintenance activities.

The use of crowdsourcing recommendations in the software development domain has been extensively investigated since 2006 [Khan et al., 2019; Mao et al., 2017; Sarı et al., 2019]. Qiao et al. in [Qiao, Yan, & Shen, 2018] introduced reinforcement learning into crowdsourcing recommendations to tackle the dilemma of cold start by incorporating a "explore & exploit" strategy to improve the effectiveness of the recommender system. A multi-model technique for engineering recommender systems was proposed by [Bouzekri et al., 2019] where authors presented a standard software architecture for RSs designed particularly for critical contexts of software systems. Williams in [Williams, 2018] developed a RS utilizing dissimilar knowledge silos to suggest attack patterns based on use case descriptions in order to alert developers and stakeholders that their software system could be compromised by mimicking the attacker's attitude during the early stages of the software development process. Assisting game developers is also presented in [Machado et al., 2019] as a RS which incorporates AI to decrease work load, improve self-efficacy, and improve accuracy. As domain modelling is another important area of research, Agt-Rickauer et al. in [Agt-Rickauer et al., 2018] supported these endeavours by designing a system for automated modelling recommendations. Linking developers with relevant tasks in open calls system has been also tackled in [Mao et al., 2015] where the authors designed content-based recommendation methods to automatically tie tasks with developers.

Another thread of efforts has been undertaken to suggest software-based components to the developers [Jorro-Aragoneses et al., 2019; Portugal et al., 2018]. Fernández-García et al. in [Fernández-García et al., 2019] created a RS with the use of machine learning algorithms to predict and recommend to developers the best cross-device component-based interfaces. Further, the evaluation and selection of the software components were addressed through a methodology proposed by Jandal et al. [Jadhav and Sonar, 2011] using hybrid a knowledge-based system technique to evaluate and recommend software packages for decision makers. One of the works undertaken to construct knowledge-based recommender systems is the CROSSMINER initiative [Bagnato et al., 2017]. This was a large-scale project whereby authors automatically collected resources and components from various open-source repositories and delivered them to the developers as recommendations using a knowledge-based RS. The outcomes of this initiative are depicted in [Nguyen et al., 2018a; Nguyen et al., 2018b, 2019]. Another means of providing recommendations for the development of a RS was proposed by Jorro-Aragoneses et al. in [Jorro-Aragoneses et al., 2019] who built a system called “RecoLibry-core” that collected components from third-party systems to assist in developing recommender systems. Commercial Off-The-Shelf (COTS) components are also significant artifacts which provide developers with high level objects that meet their needs [Cechich et al., 2006]. Yanes et al. in [Yanes et al., 2017] presented a knowledge-based personalised recommender system to infer and retrieve COTS components based on domain and linguistic ontologies. However, their approach neglected several semantic-based repositories when they established the component-based and user-based profiles. Recently, Silva et al. in [Silva et al., 2020] explored patterns in the development of Intelligent Software Engineering (ISE) solutions. To do so, they analysed 42 papers, using a thematic analysis approach to understand how previous researchers had reused knowledge and applied it to solve a SE problem. They found that researchers use internal and external knowledge sources, and rely mostly on structured data to develop ISE solutions. dos Santos et al. in [dos Santos et al., 2019] presented a knowledge-based recommendation system that considers the critical factors of a given project to indicate which multi-platform application development framework is most suitable for that project. However, the scope of the given recommendation is limited to multi-platform application development.

**Evaluation of the current approaches:** Despite the ongoing efforts to tackle obstacles that face software engineers at various stages of the development lifecycle, the sophistication of the mobile app development - as an ecosystem per se - as well as the advances in smartphone devices and various development technologies, open wide the door to future research avenues. In particular, the current approaches to recommending relevant artifacts that will assist developers in their development journey are inadequate and inferior. The following points indicate the shortcomings of the current approaches and demonstrate the superiority of our proposed system: (i) A cohort of current efforts rely mainly on clustering or statistical techniques that generally ignore the semantics of artifacts [Katirtzis et al., 2018; Moreno et al., 2015; Rubei et al., 2020; Saied et al., 2015; Wang et al., 2013; Yadav and Dutta, 2019]. Incorporating semantic analytics leveraged by a domain-ontology, semantic Web technologies and linked open data can enrich the domain ontology with instances extracted from textual contents of the collected datasets. (ii) Preferencing and profiling dimension of existing approaches is insufficient or less-focussed on the embodied techniques. The proposed approach aims to implicitly and explicitly detect developers’ interests through the construction of a multi-dimensional semantic technique that frame users’ preferences based on the designated domain. (iii) Datasets collected to evaluate the current approaches were obtained mainly from few sources [B. Lin, 2018; Rubei et al., 2020]. However, the artifacts of interest can exist in heterogenous forms and can be obtained from various resources. Our proposed approach tackles this issue by designing a system that can collect, process and retrieve artifacts obtained from tutorial and Q&A websites as well as online social networks. (iv) The temporal dimension of the current systems is not effectively incorporated. Developers’ behaviours, as well as project requirements, change over time; hence, users’ profiles and preferences may require constant amendments. (v) The current proposed recommender systems do not exclusively target the domain of mobile application development. To the best of our knowledge, this is the first attempt to tackle this issue by designing a knowledge-based time-aware personalised recommender system for mobile app development. The next section will present the proposed system and its embedded components.

## 3 Proposed System Architecture

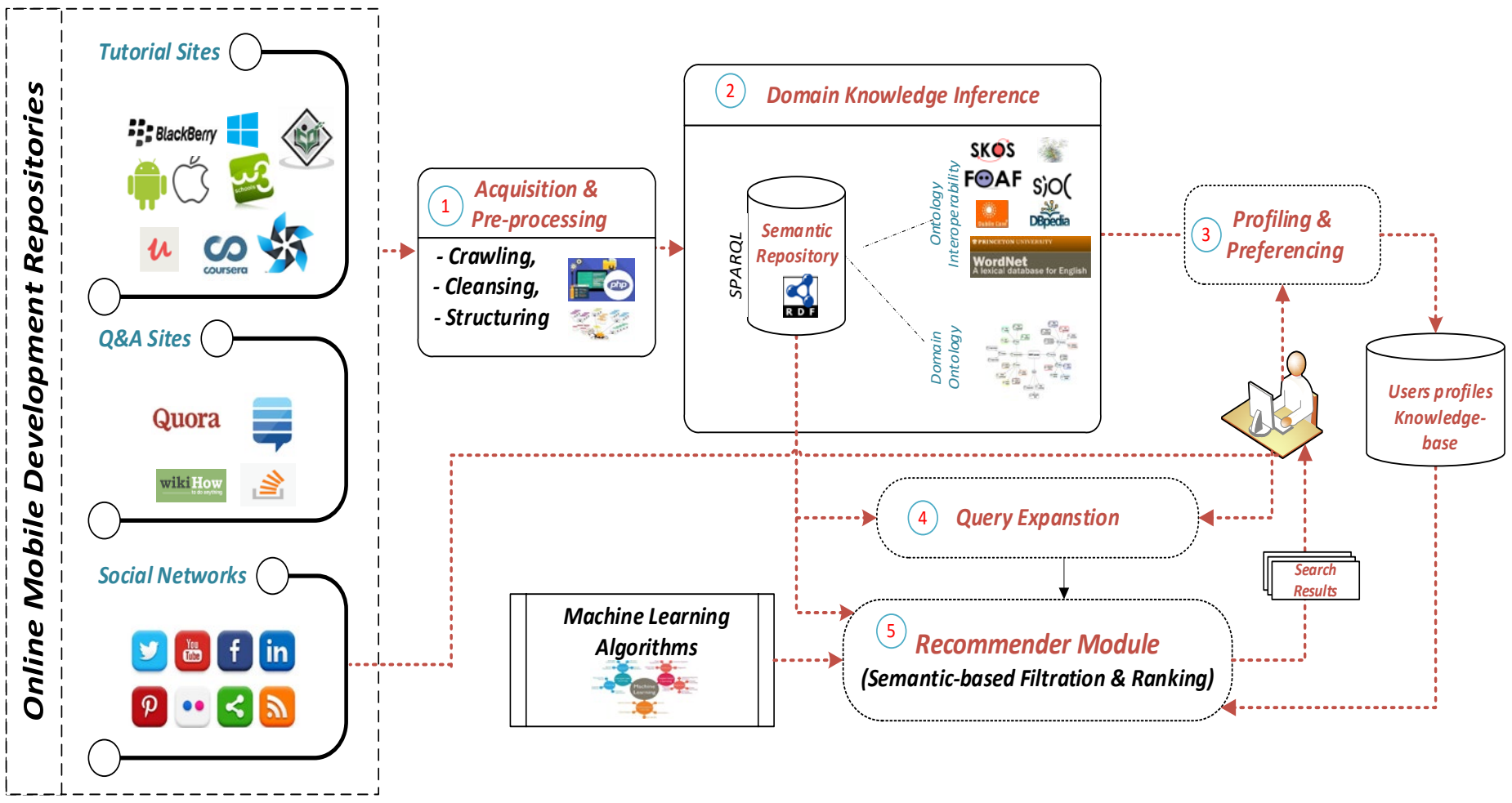

*Figure 1: Overarching System Architecture*

Figure 1 illustrates the overall architecture of the proposed system. As depicted in the diagram, the framework comprises five main stages, described below.

### 3.1 Acquisition and Pre-processing

At this stage, heterogeneous types of data are generated from different online repositories including tutorials, API documentations, Q&A repositories, to name a few. Those knowledge bases that allow APIs to access and retrieve their online content will be incorporated in this study. Moreover, data acquisition will include accessing and collecting social data pertaining to the users of the system. Several APIs will be utilized to extract batches of social data in a timely fashion.

Various pre-processing techniques are applied after the data acquisition phase: specifically, data cleansing and restructuring. During the data cleansing process, datasets are scrutinised so as to detect and deal with, for instance, corrupted, incorrect, redundant, and/or irrelevant data. Data restructuring ensures data consistency, particularly since the retrieved datasets are usually in different formats (JSON, XML, CSV, etc.). This heterogeneity of data types requires the initiation of a data structuring process to consolidate the various datasets obtained from different data islands into a single dataset that is coherent and integrated. This structured dataset facilitates management and analysis in subsequent stages of data analytics.

### 3.2 Domain Knowledge Inference

This is the "semantic kitchen" where mobile development domain ontology will be constructed to provide formal representation of the designated knowledge by identifying all the related concepts within the domain and the relationships between them. This domain ontology will provide a common shared vocabulary to be used for modelling the domain with all embodied properties and relationships.

Moreover, ontologies and various semantic web technologies and repositories are used to infer implicit knowledge from textual content as well as to model and represent the knowledge. In particular, this module will embody the ontology and semantic data interlinking techniques which facilitate the interoperability of information. The interlinking and enrichment process incorporates dissimilar vocabularies and Linked Open Data repositories such as Friend-of-a-Friend (FOAF) [Brickley and Libby, 2014], Dublin Core (DC) [Weibel, 1998], Simple Knowledge Organization System (SKOS) [Isaac and Summers, 2009; Miles and Bechhofer, 2009], Semantically-Interlinked Online Communities (SIOC) [Breslin, 2005] will be used to enrich the semantic description of resources obtained from the crawled datasets using an annotation component. In addition to ontology and vocabulary reuse, interlinking includes the semantic relationship between similar entities stored in other datasets.

The main objective of this module is twofold: to build the domain ontology for mobile app development, and to use the incorporated lightweight ontologies to enrich this domain ontology with specific semantic conceptual representations of entities obtained from the collected datasets. The second objective is to assist the process of user profiling and preferencing by means of detecting their domains of knowledge and interest obtained from examining and semantically enriching their social data content. Figure 2 shows a collection of Twitter messages about mobile app development, in particular the MAD methodologies and jobs being offered.

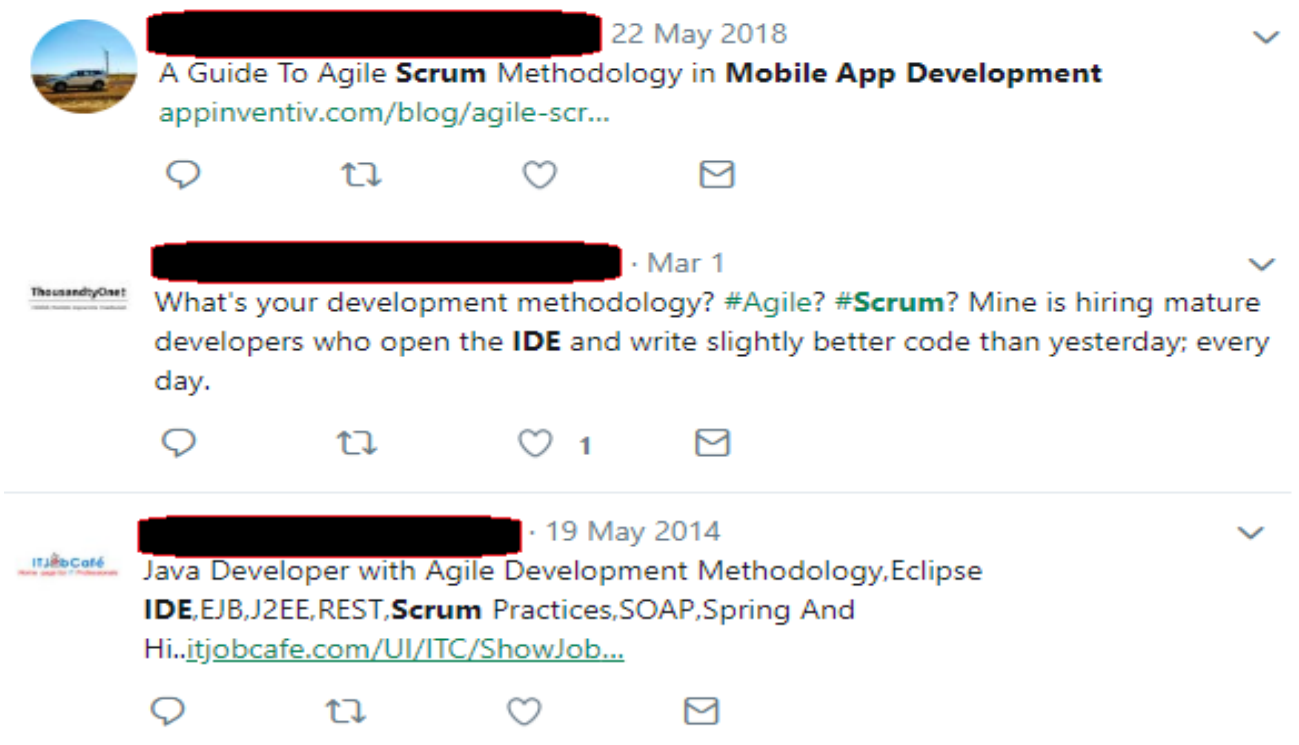

*Figure 2. Tweets Mobile App Development*

If we have an ontology as depicted in Figure 3, these tweets can be annotated and populated in the depicted ontology. Then, if the developer is looking for tutorials on MAD methodologies, then s/he would retrieve tweet #1 and tweet #3. Also, if one is looking for a job in a MAD project, s/he would receive only tweet #2, and not everything else associated with MAD. In order to facilitate this, we will need a set of rules. For example, the following rule "if there are certain instance(s) of class Tutorial e.g. scrumTutorial, then it is a tutorial of MAD" can be utilised to obtain tutorial-related concepts. Also, the following rule "if there are certain instance(s) of classes country, time e.g. month, and ticket, then it is an event" can be utilised to obtain event-related concepts. Building an ontology for MAD not only assists with the process of extracting related concepts from social media or other repositories: it will establish the necessary ground for building the recommender system and for user profiling, as discussed in the next section.

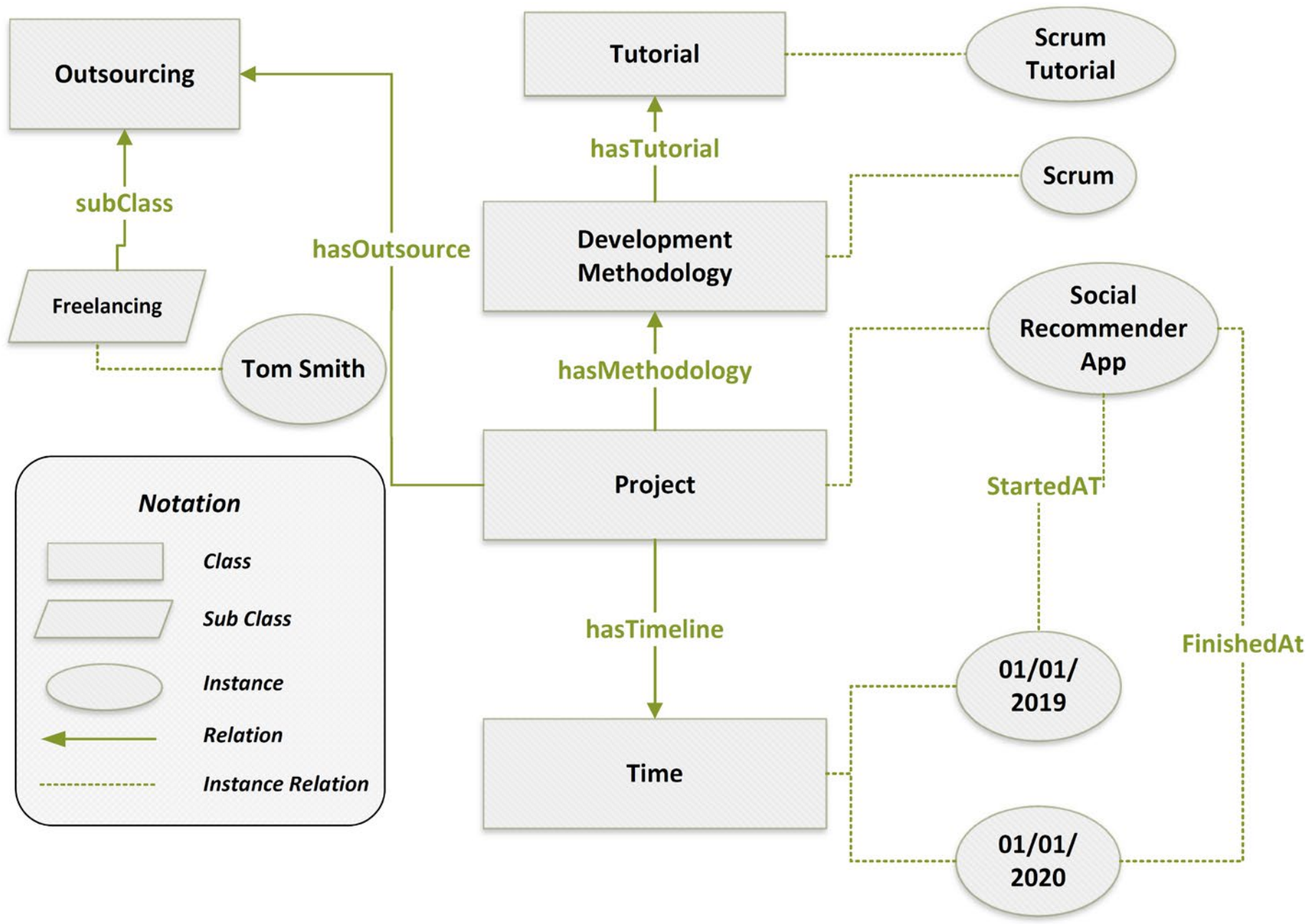

*Figure 3: Ontology representation*

### 3.3 Profiling and Preferencing

In this stage, a profile is established for each developer based on his/her interests and domain(s) of knowledge observed over a long period. The profiling and preferencing process will be designed to implicitly and explicitly detect developers' interests through their information-seeking and behaviour. To achieve this objective, users/developers will provide their user ids for Online Social Networks (OSNs) platform and/or complete a form that will be used to frame users' interests and requirements. In particular, the explicit profiling will be achieved by requesting that the user complete a designated online form to obtain the factual and available information about their domain preferences and their current projects. However, people usually try to avoid this approach either because they are not willing to disclose their personal information or because they find it tedious. Therefore, the proposal described in this paper is intended to overcome this issue by providing a hybrid approach to obtain user profiles [Poo et al., 2003]. Hence, users' interests and preferences will be inferred by analysing data contained in their OSNs, and subsequently constructing an overarching approach to obtain better user profiling which will help the recommender system to obtain tailored and personalised results for each user.

Amongst several attempts to establish techniques for user profiling and preferencing, this study has adopted and improved the multi-dimensional semantic technique presented by [Yanes et al., 2015] to model users' profiles. This model is an extension of the multi-dimensional user profiling model depicted in [Bouzeghoub and Kostadinov, 2005]. As illustrated in Figure 4, the model comprises a set of dimensions that are used to frame the user's profile. These dimensions are described briefly below.

1) *Personal Data Dimension*: this refers to the set of user's attributes which frame the user identity. This dimension can be detailed to examine various personal and demographic data although, in our context, we are more interested in a few attributes that will help to build the model. In particular, we focus primarily on information such as the developer's age, location, job title, years of experience, social media user ids, etc.

2) *Domain of Interest Dimension*: this dimension is included so as to provide specific insights into users' domains of interest. Relevant attributes will be determined using an explicit approach (i.e. online filling-up forms) and/or an implicit approach (social networks content). Items in this dimension include information about the development domain in which the developer specialises (health, business, education, misc., etc.). Also captured data pertaining to the developer's preferred app development methods (Native, Hybrid and Cross Mobile App). This dimension also includes software development methodologies (Waterfall, SCRUM, Spiral, Extreme, etc.) and software repository hosting services which are those online facilities that provide file archiving services to the affiliated members, enabling developers to store and manage their source codes online (GitHub, Buddy, AzureDevOps, etc.). The developer's domain of interest also comprises other items such as preferred programming languages, preferred IDEs, etc.

3) *Software Project Dimension*: the purpose of having this dimension is to gather any available information about the user's current software development project. This entails a set of procedures describing the software project which includes: functional and non-functional requirements, IDEs, modelling types (domain, design, etc.), programming paradigm (Object-oriented, reactive programming, component-based software engineering, etc.), front-end and back-end development tools (UI design tools, SDKs, cross-platform support, etc.), etc.

4) *Development Environment Dimension*: this dimension addresses various aspects of the environment and facility provided to develop a mobile software application in general (computer-assisted software environment). It includes aspects such as infrastructure, back-end servers (data services, authentication-authorization, integration, APIs, etc.), testing tools, debugging and troubleshooting tools for developing, testing and debugging an application or program, etc. This dimension differs from the Software Project dimension in that the development environment dimension provides a generic description of the working environment and facilities available and does not designate any specific mobile software project.

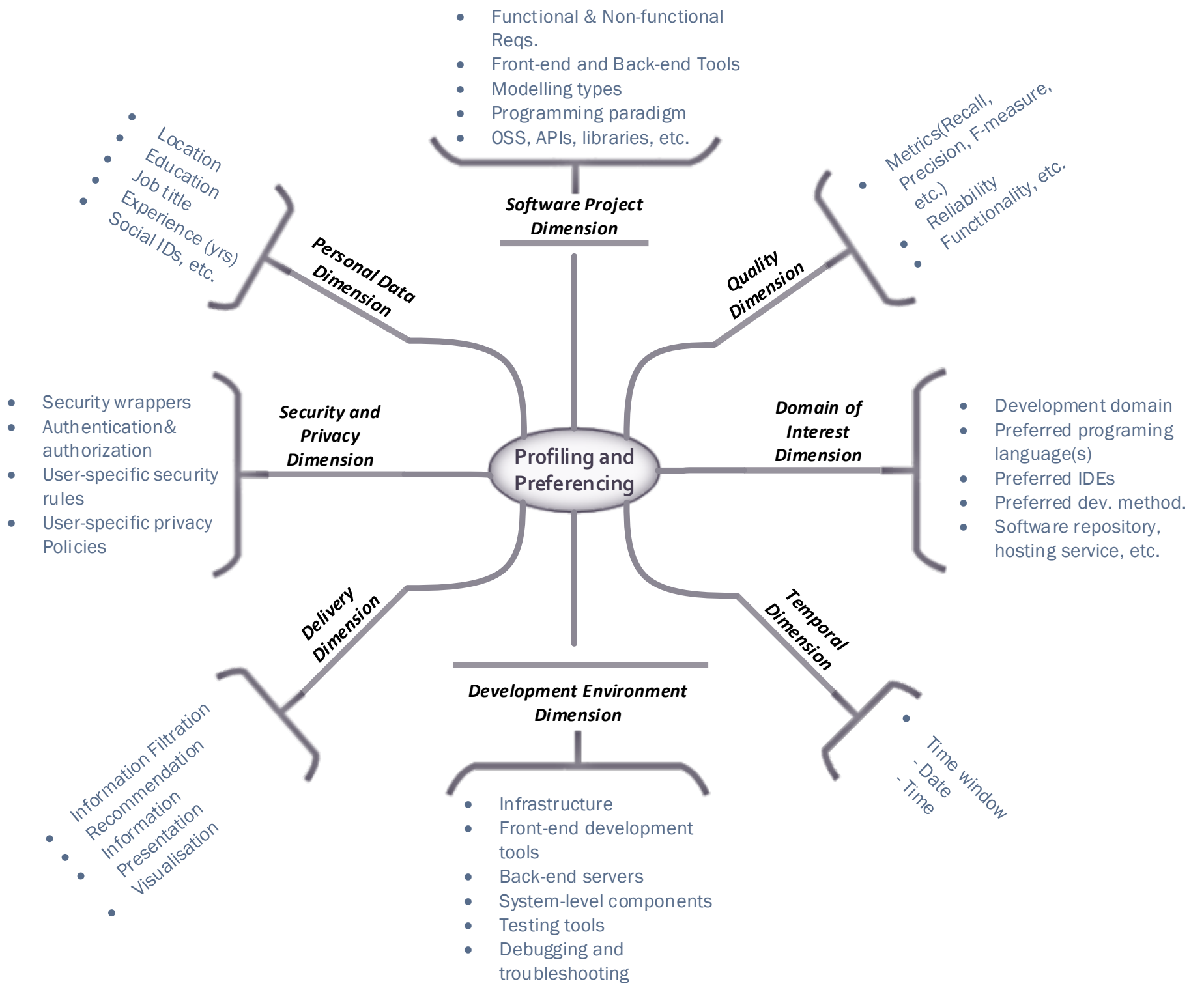

*Figure 4: Time-aware multidimensional user modelling*

5) *Security and Privacy Dimension*: this is mainly to ensure personal privacy and security by indicating all security rules and privacy policies that are established to achieve this objective. This is crucial particularly as many business firms nowadays are encouraging BYOD (Bring Your Own Device), where employees can conduct business activities using their own smartphone devices. Therefore, developers should ensure the security and privacy of users by means of app security wrapping and encryption techniques. Also, the intended approach will ensure that the identity of the developers will remain undisclosed. This also applies to details regarding the ongoing software development project and other confidential aspects.

6) *Temporal Dimension*: it is inevitable that the aforementioned dimensions will change contextually and temporally. For instance, users' interest(s) may change, and their knowledge generally evolves over time [Abu-Salih, 2018; Abu-Salih et al., 2020; Abu-Salih et al., 2018; Abu-Salih et al., 2018; Chan et al., 2018;

Meneghello et al., 2020; Nabipourshiri et al., 2018; Wongthongtham et al., 2018; Wongthongtham and Abu-Salih, 2018]. Hence, a Profiling and Preferencing module will ensure the regular updating of the user's dimensions by regularly collecting and analysing developers' social data content, and by the explicit feedback and updates obtained from the developer. This will provide up-to-date information about the user's behaviour on those platforms and will reflect the recommender system as well. This strategy will help to provide more efficient recommendation results and ensures the accuracy of temporally-updated information.

7) *Delivery Dimension*: this dimension is intended to provide the mechanism for how and what information will be delivered to the users which will be essentially the browsing and search results of the information filtration and recommender system, and how these results will be displayed to the user. In other words, the delivery dimension consists of: information filtration, recommendations, information presentation (visualisation), etc.

8) *Quality Dimension*: this dimension addresses and measures the quality of the user profiling and preferencing approach and indicates how well the model satisfies the user's requirements. This dimension will be framed by several evaluation metrics which are formulated to measure the effectiveness, functionality and reliability of the proposed profiling technique.

### 3.4 Query Expansion (QE)

QE is applied in order to improve information retrieval systems. The process involves the augmentation of a user's query with more terms, thereby obtaining better retrieval results. This process is usually automatic or interactive (semi-automatic). In the Automatic Query Expansion (AQE) approach, the developed system is responsible for selecting and augmenting a query with new terms, which differs from the interactive QE approach which infers potential terms and leaves the task of query augmentation to the user [Vechtomova, 2009]. In terms of semantic-based approaches, semantic QE can be classified into three main categories; linguistic-based, ontology-based and hybrid approaches [Raza et al., 2019]. Linguistic-based QE approaches are those techniques that generate senses of terms from thesauruses and linguistic repositories. Ontology-based approaches derive the new extended terms by semantically mapping knowledge represented in term of classes (concepts) properties and relationships as depicted in domain ontologies. Hybrid QE seeks to benefit from the two aforementioned techniques to provide a consolidated list of new terms to enhance a user's query.

This study adopts the AQE mechanism where the user's query will be augmented automatically with the help of a hybrid semantic-based QE approach. Therefore, ontologies will be used to capture domain knowledge inferred from the query, and to enrich the semantics of its textual content by providing explicit conceptual representation of entities identified in the query. Further, we make use of WordNet which is a lexical vocabulary constructed mainly to establish relationships between terms through Synsets. Synsets (or synonyms) are sets of interconnected words, terms or phrases which refer to the same semantic meaning. For instance, the words "programming, programing, computer programming, computer programing" all refer to the same semantic concept of "programing".

### 3.5 Recommendation and Information Filtration

Information filtration is the process of sieving out and delivering the right personalised information to the user. This is in fact the most imperative task of any typical Information Retrieval (IR) system. However, this is not a conventional task, particularly with the increasing proliferation of big data which makes it more difficult to collect, process, analyse and filter information. Therefore, IR systems should be designed to provide personalised services to the designated users, and to be able to effectively answer a user's query with a minimum number of mismatches.

Our aim here is to design a consolidated system that enables developers to browse and navigate through an enriched and customised catalogue of technical specifications, the latest industry updates, coding snippets, tutorials, Q&A, and a plethora of other artifacts personalised and tailored for the affiliated users. Furthermore, the intended approach will allow developers to search the knowledge bases and retrieve appropriate results that match their preferences, profile criteria, working environment and software projects specifications. This section will shed light on two crucial tasks of the proposed system.

**Task 1. Personalised IR Filtration**: one of the main objectives of this research is to build a system which allows a developer/user to search through a comprehensive list of artifacts and retrieving a list of those with the high relevancy and minimum mismatches in a ranked order style. To achieve this objective, a developer needs to submit to the system his\her query comprising a set of keywords. Then, a semantic-based query expansion technique is applied to these keywords incorporating the user's profile and his\her preferences, obtaining a consolidated set of new keywords that are added to the query, thereby expanding it. The expanded query will pass through the filtration module which is responsible to examine and retrieve a potential list of highly relevant artifacts. This curated list of artifacts will be ranked by means of the similarity ranking module. Similarity ranking will be attained by incorporating the Vector Space Model (VSM) [Salton and Yang, 1975]. VSM is a term-weighting scheme used in IR where the retrieved documents are sorted according to their degree of relevance. In VSM, a document is commonly represented by a vector of index terms exported from the document's textual content. Those index terms are associated with their computed weights indicating the significance of the index terms in the document itself and within the entire corpus. Likewise, a query is modelled to a list of index terms and weights that show the importance of each index term in the query. Cosine similarity is a core technique of VSM that is used to compute the similarity between two vectors (a document and a query). This is done by calculating the cosine value of the angle between vectors in order to find those documents most relevant to the query; the smaller the angle, the greater is the similarity between the document and the query. Cosine similarity relies on the theoretical notion of Term Frequency-Inverse Document Frequency (TF-IDF). TF-IDF measures the importance or significance of a term to a certain document within a corpus of documents, and comprises standard notions which formulate its structure. Term Frequency (TF) is used to compute the number of times that a term appears in a document. TF indicates the importance of the term in the document. Document Frequency (DF) is a statistical measure to evaluate the importance of a term to a document in a corpus of texts [Rajaraman and Ullman, 2011]. Inverse Document Frequency (IDF) is a discriminating measure for a term in the text collection. It was proposed as a cornerstone of term weighting, and is a core component of TF_IDF [Sparck Jones, 1972]. It is used as a discriminating measure to indicate the term's importance in a certain document(s) [Robertson and Sparck-Jones, 1976]. TF_IDF combines the definitions of TF (the importance of each index term in the document) and IDF (the importance of the index term in the text collection), to produce a composite weight for each term in each document. It assigns to a word t a weight in document d that is: (i) highest when t occurs many times within a few number of documents; (ii) lower when the term t occurs fewer times in a document d, or occurs in many documents; and (ii) lowest when the term t exists all documents.

In this research, the heuristic aspect will be incorporated into this model where the collected artifacts will be transformed to vectors embodying index terms mainly extracted from the textual content of these artifacts. Also, the expanded user query will be represented as a vector of index terms. Weights will be calculated for each term using the TF-IDF technique, and cosine similarity will be computed using the following formula:

$$similarity(A,B) = \frac{A \cdot B}{\|A\| \times \|B\|} = \frac{\sum_{i=1}^{n} A_i \times B_i}{\sqrt{\sum_{i=1}^{n} A_i^2} \times \sqrt{\sum_{i=1}^{n} B_i^2}}$$

where A and B are vectors represent the term frequency vectors of an artifact and the query. The resultant similarity value should range between -1 indicating no similarity, to 1 denoting that components of both the artifact and the query are identical, while intermediate values show certain levels of similarity or dissimilarity. The artifacts retrieved from the cosine similarity technique will also be automatically scrutinised to exclude those which are not related to the user's domains of interest; thus, another filtration is conducted to find matches with the user's interests obtained from the explicit or implicit approaches as described in the previous subsection. Subsequently, those artifacts which relate to the recently identified user's domains of interest will be assigned higher weights than those with less correlation.

**Task 2. Recommender System**: the intended RS will be also able to provide domain-based recommendations to users' using domain ontology and knowledge bases as an alternative to the conventional collaborative filtering and content-based RSs. An ontology-based RS is selected because knowledge-based recommender systems in general have proven their ability to address both cold-start and rating scarcity dilemmas, and can therefore be hybridised with other recommendation methods [Tarus et al., 2018]. Further, domain knowledge bases frame the recommendations so as to augment the user-resource matching, thereby providing consolidated personalised recommendations [George and Lal, 2019]. Therefore, another tool will be designed that enables the user to not only search for artifacts, but also to browse them. Domain-based semantic similarity will be applied to show the most relevant artifacts to the user, taking into consideration both the user's profile and the software project specifications. To achieve this task, various machine learning techniques will be applied to classify and predict the most relevant artifacts for the user. Examples of these algorithms include; k-nearest neighbours classifier, tree-based classifiers, multi-class logistic regression, and the SVM classifier, to name a few.

The aforementioned tasks will be attained by ensuring that user profiling and preferencing are updated regularly. The temporal dimension is crucial when designing and implementing a RS, although it is usually ignored. The developer's domain(s) of interest and expertise evolve over time. Moreover, the domain of artifacts they are seeking changes depending on the project's specifications. Hence it is necessary to tackle this fluctuation by ensuring the regular update of the user profile.

## 4 Discussion

The ubiquity and popularity of smartphones has increased the number of mobile applications and simultaneously changed various industries at a fast pace and spurred the creation of disruptive innovations [Naaman, 2016]. This has led to fierce competition between companies who are developing mobile applications, all striving to produce mobile apps that provide fast, optimized app service together with optimal user experience. However, the task of developing an app is challenging [Ahmad et al., 2018] as it requires specialised skills and knowledge and the technical capacity to keep pace with numerous tools, platforms, development environments, IDEs and devices. Further, the mechanism followed to develop and deploy a native mobile app is different from that of a regular web app.

The aforementioned issues impede mobile app developers before and during the development process, especially when a critical technical glitch or bug occurs that needs to be addressed promptly. Therefore, developers should have access to a robust mechanism that can help them to acquire the information they need before and during the development process. Developers generally rely on online sources to address technical problems encountered throughout the app development journey [B. Lin, 2018]. However, given the plethora of online platforms that provide forums where developers can obtain expertise, appropriate tools, IDEs, platforms, environment settings, and other artifacts, it takes time for a developer to acquire the information that is related to his\her domain. Hence, it is essential to develop tailored frameworks/systems which can recommend appropriate and time-aware domain-dependent tools, IDEs, platforms, environment settings, and other artifacts to mobile application developers. Also, these frameworks should be able to regularly collect, store and filter solutions, ideas and thoughts acquired from ubiquitous knowledge bases and Q&A repositories.

This paper proposes a new recommender system framework comprising a robust set of techniques designed to provide mobile app developers with a distinctive platform to browse and search for personalised artifacts. The proposed system makes use of ontology and semantic web technology as well as machine learning techniques. The following components constitute the proposed new framework: (1) Domain knowledge inference module – This module embodies various semantic web technologies and lightweight ontologies. The main objective of this module is twofold: to build the domain ontology for mobile app development, and to use the incorporated lightweight ontologies to enrich this domain ontology with specific semantic conceptual representations of entities obtained from the collected datasets. The second objective is to assist in the process of users' profiling and preferencing by means of detecting their domains of knowledge and interest obtained by examining and semantically enriching their social data content. (2) Profiling and preferencing –

This is a new proposed time-aware multidimensional user modelling. The aim of this module is to implicitly and explicitly detect developers' interests by examining their information seeking and behaviour. This module comprises various dimensions that are extracted and used to frame the user's profile. (3) Query expansion – This component aims to improve and enhance the retrieved results of the recommender system by semantically augmenting a user's query. This component makes use of domain and lightweight ontologies to enrich the textual content of the user's query, thereby expanding the query to match and retrieve further related results. (4) Recommendation and information filtration – This is the core module of the proposed system. It makes use of the aforementioned components to provide personalised services to the designated users and to answer a user's query. In particular, this essential component is responsible for undertaking two crucial tasks of the anticipated system. These are: (i) personalised information filtration that incorporates theoretically-proven IR techniques which are used to examine and retrieve a list of potentially highly relevant and ranked artifacts; and (ii) provision of domain-based time-aware recommendations to users by means of a recommender system that incorporates domain ontology and knowledge.

## 5 Conclusions and Future Work

This paper proposes a new tailored framework that is designed to support mobile app developers in their app development process. We offer a consolidated and overarching system which is able to provide personalised services and can recommend appropriate tools, IDEs, platforms, environment settings, and other artifacts from disparate online resources. The proposed system is designed to regularly collect, store and filter solutions, ideas and thoughts acquired from ubiquitous knowledge bases and Q&A repositories. In particular, the new proposed recommender system comprises the following features (i) a domain knowledge inference module that includes ontologies and various semantic web technologies and repositories; (ii) profiling and preferencing: a new time-aware multidimensional user modelling; (iii) query expansion to improve and enhance the retrieved results by semantically augmenting users' query; and (iv) recommendation and information filtration function that provides personalised services to the designated users and can answer a user's query with minimal mismatches.

In future research, we intend to develop and extend the proposed system with all embedded modules. The research will comprise the following specific tasks:

- The domain knowledge inference module will be developed and various related ontologies and linked open-data repositories will be selected and incorporated.
- The user's profiling and preferencing model will be extended and transformed to a domain ontology to factually and explicitly depict concepts and relationships representing the mobile app developers' profiles and their domain preferences.
- Query expansion is an important module which will be further scrutinised to ensure the application, enhancement and combination of state-of-the-art statistics-based techniques (such as word embedding) with semantic-based technologies.

- The intended recommender system and information filtration module will be designed and implemented to interlink the aforementioned components and also by utilizing state-of-the-art machine learning techniques.
- Another significant direction of research is to conduct sentiment analysis of the mobile application development domain in order to identify the sentiment polarity of MAD-related texts [Bin Lin et al., 2018; Wrobel, 2020]. For example, sentiment analysis can be used to scrutinize inferior artifacts that, for instance, receive negative sentiments due to the poor performance of a particular product. This can alleviate the information overload and deliver verified and positively-recommended artifacts.
- The technical interests of developers as well as the project requirements vary over time; hence, the behaviour and preferences of developers may be subject to ongoing changes. Therefore, the proposed system will integrate the temporal factor, and concept drift techniques [Lu et al., 2018] will be incorporated in the recommender system.
- In the mobile app development domain, and software engineering in general, the overall development lifecycle includes a number of stakeholders who are involved in the project in one or more phases. Therefore, the recommender system should take into account the perspectives of all parties involved (e.g. development team, project manager, executives, customer, etc.). Hence, the multistakeholder recommender system [Abdollahpouri et al., 2020] will be investigated and incorporated to address and balance the needs of multiple stakeholders in the proposed recommender systems.
- We aim to develop, refine and implement a design in order to verify and validate the proposed system and its effectiveness.

## References

[Abdollahpouri et al., 2020] Abdollahpouri, H., Adomavicius, G., Burke, R., Guy, I., Jannach, D., Kamishima, T.: Multistakeholder recommendation: Survey and research directions. User Modeling and User-Adapted Interaction, Vol. 30, No. 1 (2020), pp. 127–158.

[Abdollahpouri et al., 2020] Abdollahpouri, H., Adomavicius, G., Burke, R., Guy, I., Jannach, D., Kamishima, T., Pizzato, L.: Multistakeholder recommendation: Survey and research directions. User Modeling and User-Adapted Interaction, Vol. 30, No. 1 (2020), pp. 127–158.

[Abu-Salih 2018] Abu-Salih, B.: Trustworthiness in Social Big Data Incorporating Semantic Analysis, Machine Learning and Distributed Data Processing. Curtin University (2018).

[Abu-Salih et al., 2020] Abu-Salih, B., Al-Tawil, M., Aljarah, I., Faris, H., Wongthongtham, P.: Relational Learning Analysis of Social Politics using Knowledge Graph Embedding (arXiv preprint arXiv: 2006.01626) (2020).

[Abu-Salih et al., 2020] Abu-Salih, B., Chan, K. Y., Al-Kadi, O., Al-Tawil, M., Wongthongtham, P., Issa, T., Albahlal, A.: Time-aware domain-based social influence prediction. Journal of Big Data, Vol. 7, No. 1 (2020). https://doi.org/10. doi: 10.1186/s40537-020-0283-3

[Abu-Salih et al., 2018] Abu-Salih, B., Wongthongtham, P., Chan, K. Y.: Twitter mining for ontology-based domain discovery incorporating machine learning. Journal of Knowledge Management, Vol. 22, No. 5 (2018), pp. 949–981. https://doi.org/doi: 10.1108/jkm-11-2016-0489

[Abu-Salih et al., 2018] Abu-Salih, B., Wongthongtham, P., Chan, K. Y., Zhu, D.: CredSaT: Credibility ranking of users in big social data incorporating semantic analysis and temporal factor. Journal of Information Science, Vol. 45, No. 2 (2018), pp. 259–280. https://doi.org/doi: 10.1177/0165551518790424

[Agt-Rickauer et al., 2018] Agt-Rickauer, H., Kutsche, R.-D., Sack, H.: DoMoRe-A Recommender System for Domain Modeling. Paper presented at the MODELSWARD (2018).

[Ahmad et al., 2018] Ahmad, A., Li, K., Feng, C., Asim, S. M., Yousif, A., Ge, S.: An empirical study of investigating mobile applications development challenges. IEEE Access, Vol. 6 (2018), pp. 17711–17728.

[Bagnato et al., 2017] Bagnato, A., Barmpis, K., Bessis, N., Cabrera-Diego, L. A., Di Rocco, J., Di Ruscio, D., Krief, P.: Developer-centric knowledge mining from large open-source software repositories (crossminer). Paper presented at the Federation of International Conferences on Software Technologies: Applications and Foundations (2017).

[Beyer 2015] Beyer, S.: DIETs: recommender systems for mobile API developers. Paper presented at the 2015 IEEE/ACM 37th IEEE International Conference on Software Engineering (2015).

[Bouzeghoub and Kostadinov 2005] Bouzeghoub, M., Kostadinov, D.: Personnalisation de l'information: aperçu de l'état de l'art et définition d'un modèle flexible de profils. CORIA, Vol. 5 (2005), pp. 201–218.

[Bouzekri et al., 2019] Bouzekri, E., Canny, A., Fayollas, C., Martinie, C., Palanque, P., Barboni, E., Gris, C.: Engineering issues related to the development of a recommender system in a critical context: Application to interactive cockpits. International Journal of Human-Computer Studies, Vol. 121 (2019), pp. 122–141. https://doi.org/doi:https://doi.org/10.1016/j.ijhcs.2018.05.001

[Breslin 2005] Breslin, J. G.: Towards semantically-interlinked online communities. Springer Berlin Heidelberg (2005).

[Brickley and Libby 2010] Brickley, D., Libby, M.: FOAF vocabulary specification 0.98; (2010).

[Chan et al., 2018] Chan, K. Y., Kwong, C. K., Wongthongtham, P., Jiang, H., Fung, C. K. Y., Abu-Salih, B., Jain, P.: Affective design using machine learning: a survey and its prospect of conjoining big data. International Journal of Computer Integrated Manufacturing, Vol. 1, No. 19 (2018). https://doi.org/doi: 10.1080/0951192x.2018.1526412

[Charkaoui and Adraoui 2014] Charkaoui, S., Adraoui, Z.: Cross-platform mobile development approaches. Paper presented at the 2014 Third IEEE International Colloquium in Information Science and Technology (CIST) (2014).

[Christensen and Schiaffino 2011] Christensen, I. A., Schiaffino, S.: Entertainment recommender systems for group of users. Expert Systems with Applications, Vol. 38, No. 11 (2011), pp. 14127–14135.

[dos Santos et al., 2019] dos Santos, D. S., Nunes, H. D., Macedo, H. T., Neto, A. C.: Recommendation System for Cross-Platform Mobile Development Framework. Paper presented at the Proceedings of the XV Brazilian Symposium on Information Systems (2019).

[Fernández-García et al., 2019] Fernández-García, A. J., Iribarne, L., Corral, A., Criado, J., Wang, J. Z.: A recommender system for component-based applications using machine learning techniques. Knowledge-Based Systems, Vol. 164 (2019), pp. 68–84.

[George and Lal 2019] George, G., Lal, A. M.: Review of ontology-based recommender systems in e-learning. Computers & Education, Vol. 103642 (2019).

[Ibarra and Rodriguez 2019] Ibarra, R., Rodriguez, G.: SoTesTeR: Software Testing Techniques Recommender System Using a Collaborative Approach. Cham (2019).

[Isaac and Summers 2009] Isaac, A., Summers, E.: Skos simple knowledge organization system. Primer, World Wide Web Consortium, Vol. 3 (2009), p. 7.

[Jadhav and Sonar 2011] Jadhav, A. S., Sonar, R. M.: Framework for evaluation and selection of the software packages: A hybrid knowledge-based system approach. Journal of Systems and Software, Vol. 84, No. 8 (2011), pp. 1394–1407.

[Jason Wong et al., 2018] Jason Wong, V. B., Clark, W., Leow, A., Resnick, M., Driver, M., Revang, M., Dunie, R.: Technology Insight for Multiexperience Development Platforms. (2018). Retrieved from https://www.gartner.com/document/3859463

[Jorro-Aragoneses et al., 2019] Jorro-Aragoneses, J. L., Recio-García, J. A., Díaz-Agudo, B., Jimenez-Díaz, G.: RecoLibry-core: A component-based framework for building recommender systems. Knowledge-Based Systems, Vol. 104854 (2019).

[Katirtzis et al., 2018] Katirtzis, N., Diamantopoulos, T., Sutton, C. A.: Summarizing Software API Usage Examples Using Clustering Techniques. Paper presented at the FASE (2018).

[Khan et al., 2019] Khan, J. A., Liu, L., Wen, L., Ali, R.: Crowd Intelligence in Requirements Engineering: Current Status and Future Directions. Paper presented at the International Working Conference on Requirements Engineering: Foundation for Software Quality (2019).

[Lin 2018] Lin, B.: Crowdsourced Software Development and Maintenance. Paper presented at the 2018 IEEE/ACM 40th International Conference on Software Engineering: Companion (ICSE-Companion) (Vol. 27) (2018).

[Lin et al., 2018] Lin, B., Zampetti, F., Bavota, G., Di Penta, M., Lanza, M., Oliveto, R.: Sentiment analysis for software engineering: How far can we go? Paper presented at the Proceedings of the 40th International Conference on Software Engineering (2018).

[Lu et al., 2018] Lu, J., Liu, A., Dong, F., Gu, F., Gama, J., Zhang, G.: Learning under concept drift: A review. IEEE Transactions on Knowledge and Data Engineering, Vol. 31, No. 12 (2018), pp. 2346–2363.

[Machado et al., 2019] Machado, T., Gopstein, D., Nov, O., Wang, A., Nealen, A., Togelius, J.: Evaluation of a Recommender System for Assisting Novice Game Designers. (arXiv preprint arXiv: 1908.04629) (2019).

[Majchrzak et al., 2018] Majchrzak, T. A., Biørn-Hansen, A., Grønli, T.-M.: Progressive web apps: the definite approach to Cross-Platform development? (2018).

[Mao et al., 2017] Mao, K., Capra, L., Harman, M., Jia, Y.: A survey of the use of crowdsourcing in software engineering. Journal of Systems and Software, Vol. 126 (2017), pp. 57–84.

[Mao et al., 2015] Mao, K., Yang, Y., Wang, Q., Jia, Y., Harman, M.: Developer recommendation for crowdsourced software development tasks. Paper presented at the 2015 IEEE Symposium on Service-Oriented System Engineering, (2015).

[Meneghello et al., 2020] Meneghello, J., Thompson, N., Lee, K., Wong, K. W., Abu-Salih, B.: Unlocking Social Media and User Generated Content as a Data Source for Knowledge Management. International Journal of Knowledge Management (IJKM), Vol. 16, No. 1 (2020), pp. 101–122.

[Miles and Bechhofer 2009] Miles, A., Bechhofer, S.: SKOS simple knowledge organization system reference (2009).

[Muhairat 2020] Muhairat, M.: An Intelligent Recommender System Based on Association Rule Analysis for Requirement Engineering. J. UCS, Vol. 26, No. 1 (2020), pp. 33–49.

[Moreno et al., 2015] Moreno, L., Bavota, G., Di Penta, M., Oliveto, R., Marcus, A.: How can I use this method?. Paper presented at the Proceedings of the 37th International Conference on Software Engineering-Volume (2015).

[Naaman 2016] Naaman, S.: Disruptive Technologies and Local Governments: The Case of Uber in San Francisco and New York City. Cornell University (2016).

[Nabipourshiri et al., 2018] Nabipourshiri, R., Abu-Salih, B., Wongthongtham, P.: Tree-based Classification to Users' Trustworthiness in OSNs. Paper presented at the Proceedings of the 2018 10th International Conference on Computer and Automation Engineering - ICCAE. 2018, Brisbane, Australia (2018).

[Nguyen et al., 2018a] Nguyen, P. T., Di Rocco, J., Di Ruscio, D.: Knowledge-aware Recommender System for Software Development. Paper presented at the KaRS@ RecSys, (2018a).

[Nguyen et al., 2018b] Nguyen, P. T., Di Rocco, J., Di Ruscio, D.: Mining Software Repositories to Support OSS Developers: A Recommender Systems Approach. Paper presented at the IIR, (2018b).

[Nguyen et al., 2019] Nguyen, P. T., Di Rocco, J., Di Ruscio, D.: Enabling heterogeneous recommendations in OSS development: what's done and what's next in CROSSMINER. Paper presented at the Proceedings of the Evaluation and Assessment on Software Engineering (2019).

[Polatidis and Georgiadis 2013] Polatidis, N., Georgiadis, C. K.: Recommender systems: The Importance of personalization in E-business environments. International Journal of E-Entrepreneurship and Innovation (IJEEI), Vol. 4, No. 4 (2013), pp. 32–46.

[Poo, Chng and Goh 2003] Poo, D., Chng, B., Goh, J.-M.: A hybrid approach for user profiling. Paper presented at the Proceedings of the (2003).

[Portugal et al., 2018] Portugal, I., Alencar, P., Cowan, D.: The use of machine learning algorithms in recommender systems: A systematic review (2018), pp. 205–227.

[Qiao et al., 2018] Qiao, R., Yan, S., Shen, B.: A Reinforcement Learning Solution to Cold-Start Problem in Software Crowdsourcing Recommendations. Paper presented at the 2018 IEEE International Conference on Progress in Informatics and Computing (PIC) (2018), pp. 14–16.

[Rajaraman and Ullman 2011] Rajaraman, A., Ullman, J. D.: Mining of Massive Datasets; Cambridge University Press (2011).

[Raza et al., 2019] Raza, M. A., Mokhtar, R., Ahmad, N., Pasha, M., Pasha, U.: A Taxonomy and Survey of Semantic Approaches for Query Expansion. IEEE Access, Vol. 7 (2019), pp. 17823–17833.

[Resnick and Varian 1997] Resnick, P., Varian, H. R.: Recommender systems.; Communications of the ACM, Vol. 40, No. 3 (1997), pp. 56–58.

[Ricci et al., 2015] Ricci, F., Rokach, L., Shapira, B.: Recommender systems: introduction and challenges. Recommender Systems Handbook. Springer (2015), pp. 1–34.

[Robertson and Sparck-Jones 1976] Robertson, S. E., Sparck-Jones, K.: Relevance Weighting of Search Terms. Journal of the American Society for Information Science, Vol. 27, No. 3 (1976), pp. 129–146. https://doi.org/doi: DOI 10.1002/asi.4630270302.

[Robillard et al., 2009] Robillard, M., Walker, R., Zimmermann, T.: Recommendation systems for software engineering. IEEE Software, Vol. 27, No. 4 (2009), pp. 80–86.

[Rubei et al., 2020] Rubei, R., Di Sipio, C., Nguyen, P. T., Di Rocco, J., Di Ruscio, D.: PostFinder: Mining Stack Overflow posts to support software developers. Information and Software Technology, Vol. 127 (2020), p. 106367.

[Saied et al., 2015] Saied, M. A., Benomar, O., Abdeen, H., Sahraoui, H.: Mining multi-level api usage patterns. Paper presented at the 2015 IEEE 22nd International Conference on Software Analysis, Evolution, and Reengineering (SANER) (2015).

[Salton et al., 1975] Salton, G., Wong, A., Yang, C.-S.: A vector space model for automatic indexing. Communications of the ACM, Vol. 18, No. 11 (1975), pp. 613–620.

[Sarı et al., 2019] Sarı, A., Tosun, A., Alptekin, G. I.: A systematic literature review on crowdsourcing in software engineering. Journal of Systems and Software, Vol. 153 (2019), pp. 200–219.

[Silva et al., 2020] Silva, L. A. P., Chagas, J. F. S., Perkusich, M., de Sousa Neto, A. F., Albuquerque, D., Valadares, D. C. G., Perkusich, A.: On the Reuse of Knowledge to Develop Intelligent Software Engineering Solutions. Paper presented at The 32nd International Conference on Software Engineering and Knowledge Engineering, SEKE 2020, KSIR Virtual Conference Center, USA (2020), p. USA. Retrieved from https://doi.org/10.18293/SEKE2020-157

[Sparck Jones 1972] Sparck Jones, K.: A statistical interpretation of term specificity and its application in retrieval. Journal of Documentation, Vol. 28, No. 1 (1972), pp. 11–21.

[Sun et al., 2015] Sun, Z., Han, L., Huang, W., Wang, X., Zeng, X., Wang, M., Yan, H.: Recommender systems based on social networks. Journal of Systems and Software, Vol. 99 (2015), pp. 109–119.

[Tarus et al., 2018] Tarus, J. K., Niu, Z., Mustafa, G.: Knowledge-based recommendation: a review of ontology-based recommender systems for e-learning. Artificial Intelligence Review, Vol. 50, No. 1 (2018), pp. 21–48.

[Umuhoza and Brambilla 2016] Umuhoza, E., Brambilla, M.: Model driven development approaches for mobile applications: A survey. Paper presented at the International Conference on Mobile Web and Information Systems (2016).

[Valdez et al., 2016] Valdez, A. C., Ziefle, M., Verbert, K., Felfernig, A., Holzinger, A.: Recommender systems for health informatics: state-of-the-art and future perspectives. Machine Learning for Health Informatics. Springer (2016), pp. 391–414.

[Vechtomova 2009] Vechtomova, O.: Query Expansion for Information Retrieval; in L. Liu & M. T. ÖZsu (Eds.), Encyclopedia of Database Systems. Boston, MA: Springer US (2009), pp. 2254–2257.

[Wang et al., 2013] Wang, J., Dang, Y., Zhang, H., Chen, K., Xie, T., Zhang, D.: Mining succinct and high-coverage API usage patterns from source code. Paper presented at the 2013 10th Working Conference on Mining Software Repositories (MSR) (2013), pp. 319–328. https://doi.org/10.1109/msr.2013.6624045

[Weibel 1998] Weibel, S.: Dublin core metadata for resource discovery. Internet Engineering Task Force RFC, Vol. 2413 (1998), p. 222.

[Wiesner and Pfeifer 2014] Wiesner, M., Pfeifer, D.: Health recommender systems: concepts, requirements, technical basics and challenges. International Journal of Environmental Research and Public Health, Vol. 11, No. 3 (2014), pp. 2580–2607.

[Williams 2018] Williams, I.: An Ontology Based Collaborative Recommender System for Security Requirements Elicitation. Paper presented at the 2018 IEEE 26th International Requirements Engineering Conference (RE) (2018).

[Wongthongtham et al., 2018] Wongthongtham, P., Chan, K. Y., Potdar, V., Abu-Salih, B., Gaikwad, S., Jain, P.: State-of-the-Art Ontology Annotation for Personalised Teaching and Learning and Prospects for Smart Learning Recommender Based on Multiple Intelligence and Fuzzy Ontology; International Journal of Fuzzy Systems, Vol. 20, No. 4 (2018), pp. 1357–1372. https://doi.org/doi: 10.1007/s40815-018-0467-6

[Wongthongtham and Salih 2018] Wongthongtham, P., Salih, B. A.: Ontology-based approach for identifying the credibility domain in social Big Data.; Journal of Organizational Computing and Electronic Commerce, Vol. 28, No. 4 (2018), pp. 354–377.

[Wrobel 2020] Wrobel, M. R.: The Impact of Lexicon Adaptation on the Emotion Mining from Software Engineering Artifacts; IEEE Access, Vol. 8 (2020), pp. 48742–48751.

[Yadav and Dutta 2019] Yadav, D. K., Dutta, S. K.: Test Case Prioritization Using Clustering Approach for Object Oriented Software; International Journal of Information System Modeling and Design (IJISMD), Vol. 10, No. 3 (2019), pp. 92–109.

[Yanes et al., 2015] Yanes, N., Ben Sassi, S., Ben Ghezala, H.: A Multidimensional Semantic User Model for COTS Components Search Personalization. Paper presented at the The 26th; In IBIMA Conference on Innovation and Sustainable Economic Competitive Advantage: From Regional Development to Global Growth. Madrid, Spain (2015).

[Yanes et al., 2017] Yanes, N., Sassi, S. B., Ghezala, H. H. B.: Ontology-based recommender system for COTS components; Journal of Systems and Software, Vol. 132 (2017), pp. 283–297.